\documentclass[aps,preprint,pre,superscriptaddress,runinaddress]{revtex4} 

\usepackage{amssymb}

\usepackage{graphicx}

\begin{document}



\title{Classical Representation of a Quantum System at Equilibrium}



\author{James W. Dufty\footnote{Corresponding author\quad
E-mail:~\textsf{Dufty@phys.ufl.edu}}}
 \altaffiliation{Department of Physics, University of Florida, Gainesville, FL 32611}
\author{Sandipan Dutta}%
 \email{duttas@phys.ufl.edu}
\affiliation{%
 Department of Physics, University of Florida, Gainesville, FL 32611
}%
\date{\today}

\begin{abstract}
  A quantum system at equilibrium is represented by a corresponding
classical system, chosen to reproduce the thermodynamic and
structural properties. The objective is to develop a means for
exploiting strong coupling classical methods (e.g., MD, integral
equations, DFT) to describe quantum systems. The classical system
has an effective temperature, local chemical potential, and pair
interaction that are defined by requiring equivalence of the grand
potential and its functional derivatives with respect to the
external and pair potentials for the classical and quantum systems.
Practical inversion of this mapping for the classical properties is
effected via the hypernetted chain approximation, leading to
representations as functionals of the quantum pair correlation
function. As an illustration, the parameters of the classical system
 are determined approximately such
that ideal gas and weak coupling RPA limits
are preserved. 

\end{abstract}
\maketitle                   
\section{Introduction}
A simple atomic system in its liquid state is a prototypical
strongly coupled system with no small expansion parameters available
to simplify calculations. Under conditions where classical mechanics
is applicable there are a number of theoretical and computational
methods available to address such strong coupling conditions. These
include integral equation and density functional methods, and
numerical techniques such as molecular dynamics (MD) and Monte Carlo
(MC) simulation. For
quantum systems there are complementary many-body methods to address
the new features of diffraction and exchange or degeneracy effects.
Conditions of both quantum degeneracy and strong coupling pose
challenging problems with few choices for accurate computation of
properties sensitive to both features (e.g., diffusion and path
integral Monte Carlo (PIMC) simulation).

One approach to describe such quantum systems is to extend the
classical methods in a phenomenological way. For example, MD has
been used with effective potentials modified to include short range
diffraction effects \cite{Filinov}. A related approach, wave packet MD, uses a
similar variant of classical mechanics to define trajectories
guiding wave-packets. An interesting new approach is to use
classical integral equations for pair correlations
modified to include an effective potential and an effective temperature \cite%
{Dharma00}. Quantum effects are imbedded in these effective
properties, while strong correlations are generated by the classical
form of the integral equation. A surprisingly wide class of quantum
systems have been described in this way with considerable success
\cite{Dharma11}.

The objective here is to formalize the exploitation of classical
methods by defining a classical statistical mechanics whose
equilibrium properties are the same as those for an underlying
quantum system. The exact definition is given in the next section,
providing the basis for the introduction of practical
approximations. One simple realization of electron gas as a
classical system is described in the following sections. The
emphasis here is on the definition of the mapping and its approximate
evaluation. Applications will be described elsewhere.

\section{Definition of the classical system}
\label{sec2} Consider for simplicity a one component system at
equilibrium represented in the grand canonical ensemble. The grand
potential $\Omega (\beta \mid \mu,\phi )$
is proportional to the pressure and is given by%
\begin{equation}
\Omega (\beta \mid \mu,\phi )=-p(\beta \mid \mu,\phi )V=-\beta ^{-1}\ln
\sum_{N}Tr_{N}e^{-\beta \left( K+\Phi -\int d\mathbf{r}\mu (\mathbf{r})%
\widehat{n}(\mathbf{r})\right) }.  \label{2.1}
\end{equation}%
Here $Tr_{N}$ denotes a trace over properly symmetrized $N$ particle
states. The thermodynamics for the system is specified as functions
of the inverse
temperature $\beta $ and functionals of the local chemical potential $\mu (%
\mathbf{r})\equiv \mu -\phi _{ext}\left( \mathbf{r}\right) $ (an
external potential has been included for generality) and pair potential $\phi$. The
Hamiltonian $H=K+\Phi $
contains the kinetic energy $K$ and sum over pair potentials:
$\Phi =\frac{1}{2}\sum_{ij}\phi \left( q_{ij}\right) $.

A corresponding classical system is considered with Hamiltonian
$H_{c}$ in the same volume $V$ at equilibrium described by the
classical grand canonical ensemble, with inverse temperature $\beta
_{c}$ and local chemical
potential $\mu _{c}(\mathbf{r})=\mu _{c}-\phi _{c,ext}\left( \mathbf{r}%
\right) $ . The Hamiltonian has the same form except the potential
energy
functions $\phi _{c}\left( q_{ij}\right) $ and $\phi _{c,ext}\left( \mathbf{q%
}_{i}\right) $ are different. The classical grand potential is
defined in
terms of these quantities by%
\begin{equation}
\Omega _{c}(\beta _{c}\mid \mu _{c},\phi_c)=-p_{c}(\beta _{c}\mid \mu
_{c},\phi_c)V=-\frac{1}{\beta_c}\ln \sum_{N}\frac{1}{\lambda _{c}^{3N}N!}\int d\mathbf{q}_{1}..d%
\mathbf{q}_{N}e^{-\beta _{c}\left( \Phi _{c}-\int dr\mu _{c}(r)\widehat{n}%
(r)\right) }.  \label{2.3}
\end{equation}%
Here, $\lambda _{c}=\left( 2\pi \beta _{c}\hbar ^{2}/m\right)
^{1/2}$ is the thermal de Broglie wavelength associated with the
classical temperature. The integration for the partition function is
taken over the $N$ particle configuration space.

The classical system has undefined ingredients: the effective
inverse
temperature, $\beta _{c}$, the local chemical potential, $\mu _{c}(\mathbf{r}%
)$, and the pair potential for interaction among the particles,
$\phi _{c}\left( q_{ij}\right) $. A correspondence between the
classical and quantum systems is defined by expressing these
quantities as functions or functionals of $\beta $, $\mu
(\mathbf{r})$, and $\phi \left( q_{ij}\right) $. This is accomplished
by requiring the numerical equivalence of two independent
thermodynamic properties and one structural property. The first two
are chosen to be the equivalence of the grand potential and its
first functional derivative with respect to the local chemical
potential.
\begin{equation}
\Omega _{c}(\beta _{c}\mid \mu _{c},\phi_c)\equiv \Omega (\beta \mid \mu,\phi),\hspace{%
0.25in}\frac{\delta \Omega _{c}(\beta _{c}\mid \mu _{c},\phi_c)}{\delta \mu _{c}(%
\mathbf{r})}\mid _{\beta _{c},\phi _{c}}\equiv \frac{\delta \Omega
(\beta \mid \mu,\phi )}{\delta \mu (\mathbf{r})}\mid _{\beta,\phi }.
\label{2.4}
\end{equation}%
An equivalent form for these conditions can be given in terms of the
pressure and density%
\begin{equation}
p_{c}(\beta _{c}\mid \mu _{c},\phi_c)\equiv p(\beta \mid \mu,\phi ),\hspace{0.25in}n_{c}(%
\mathbf{r;}\beta _{c}\mid \mu _{c},\phi_c)\equiv n(\mathbf{r;}\beta \mid
\mu,\phi ). \label{2.5}
\end{equation}%
These two relations provide two independent relations between $\beta _{c}$%
, $\mu _{c}(\mathbf{r})$ and the physical variables $\beta $ and $\mu (%
\mathbf{r})$.  It remains to have a structural equivalence to relate
the pair potential $\phi _{c}\left( q_{ij}\right) $ to $\phi \left(
q_{ij}\right) $, which are two particle functions. This is
accomplished by equating the functional derivatives of the grand
potentials with respect to
these pair potentials%
\begin{equation}
\frac{\delta \Omega _{c}(\beta _{c}\mid \mu _{c},\phi_c)}{%
\delta \phi _{c}(\mathbf{r},\mathbf{r}^{\prime })}\mid_{\beta_c,\mu_c}=\frac{%
\delta \Omega (\beta \mid \mu,\phi )}{\delta \phi
(\mathbf{r},\mathbf{r}^{\prime })}\mid_{\beta,\mu}.  \label{2.6}
\end{equation}%
An equivalent form for this definition is the equivalence of pair
correlation functions%
\begin{equation}
g_{c}(\mathbf{r},\mathbf{r}^{\prime };\beta _{c}\mid \mu _{c},\phi_c)\equiv g(%
\mathbf{r},\mathbf{r}^{\prime };\beta \mid \mu,\phi ).  \label{2.7}
\end{equation}%

In this way the three equations of (\ref{2.5}) and (\ref{2.7})
determine, formally, the classical parameters $\beta _{c},$ $\mu
_{c},$ and $\phi
_{c}\left( q\right) $ as functions of $\beta ,$ and functionals of $\mu (%
\mathbf{r})$, and $\phi \left( q\right) $:
$\beta _{c}=\beta _{c}(\beta \mid \mu,\phi ),\hspace{0.1in}\mu _{c}=\mu _{c}(%
\mathbf{r;}\beta \mid \mu,\phi ),\hspace{0.1in}\phi _{c}=\phi _{c}\left( \mathbf{%
r,r}^{\prime };\beta \mid \mu,\phi \right)$.  This completes the
definition of the thermodynamics of a classical system that is
representative of that for a given quantum system. It is also
possible to deduce from this a classical density functional theory
that is representative of the quantum density functional theory.
This will be discussed elsewhere.

\section{Inversion of the map}
\label{sec3}
\begin{figure}[<float>]
\includegraphics[width=80mm,height=55mm]{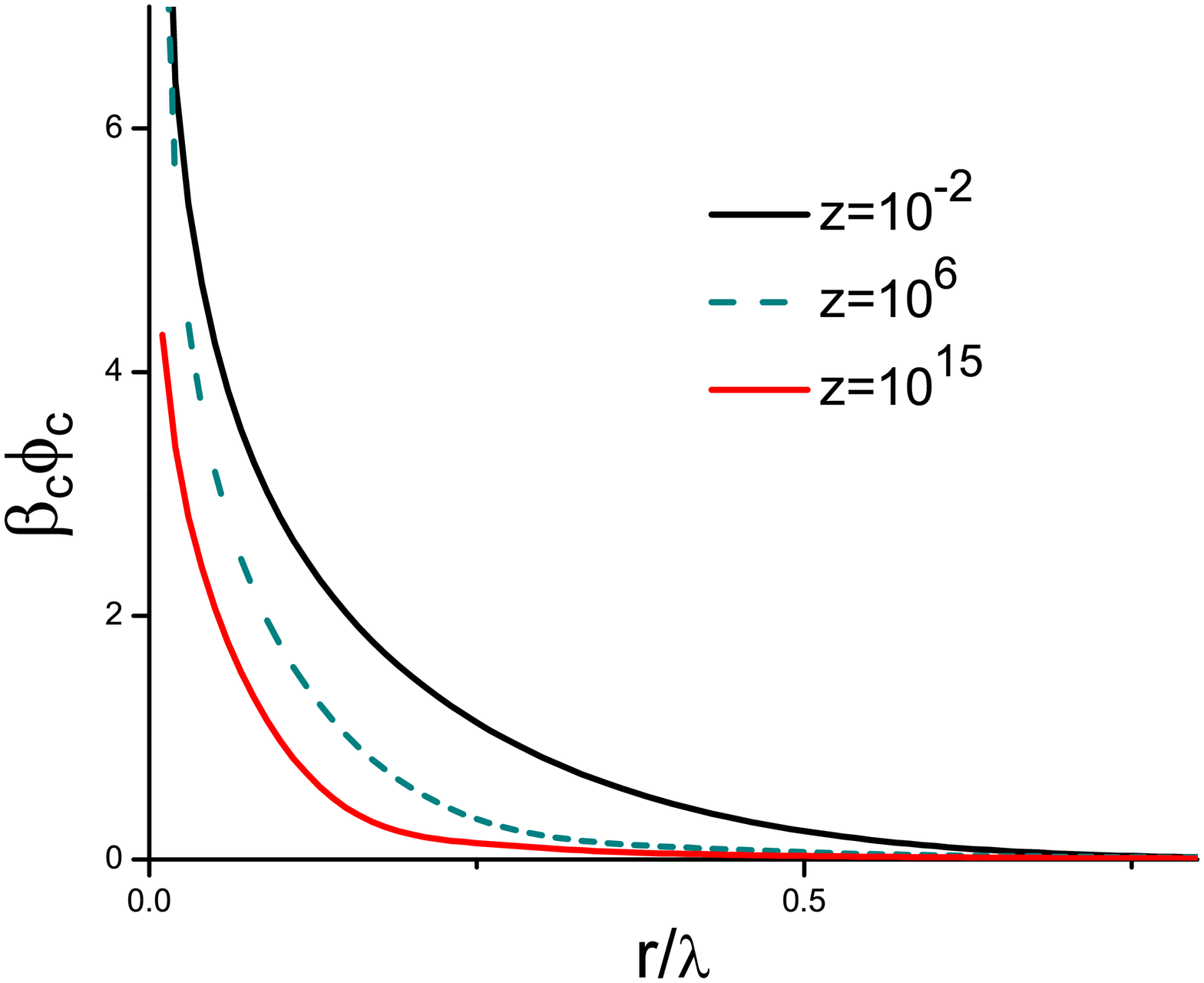}
\hfil
\includegraphics[width=80mm,height=55mm]{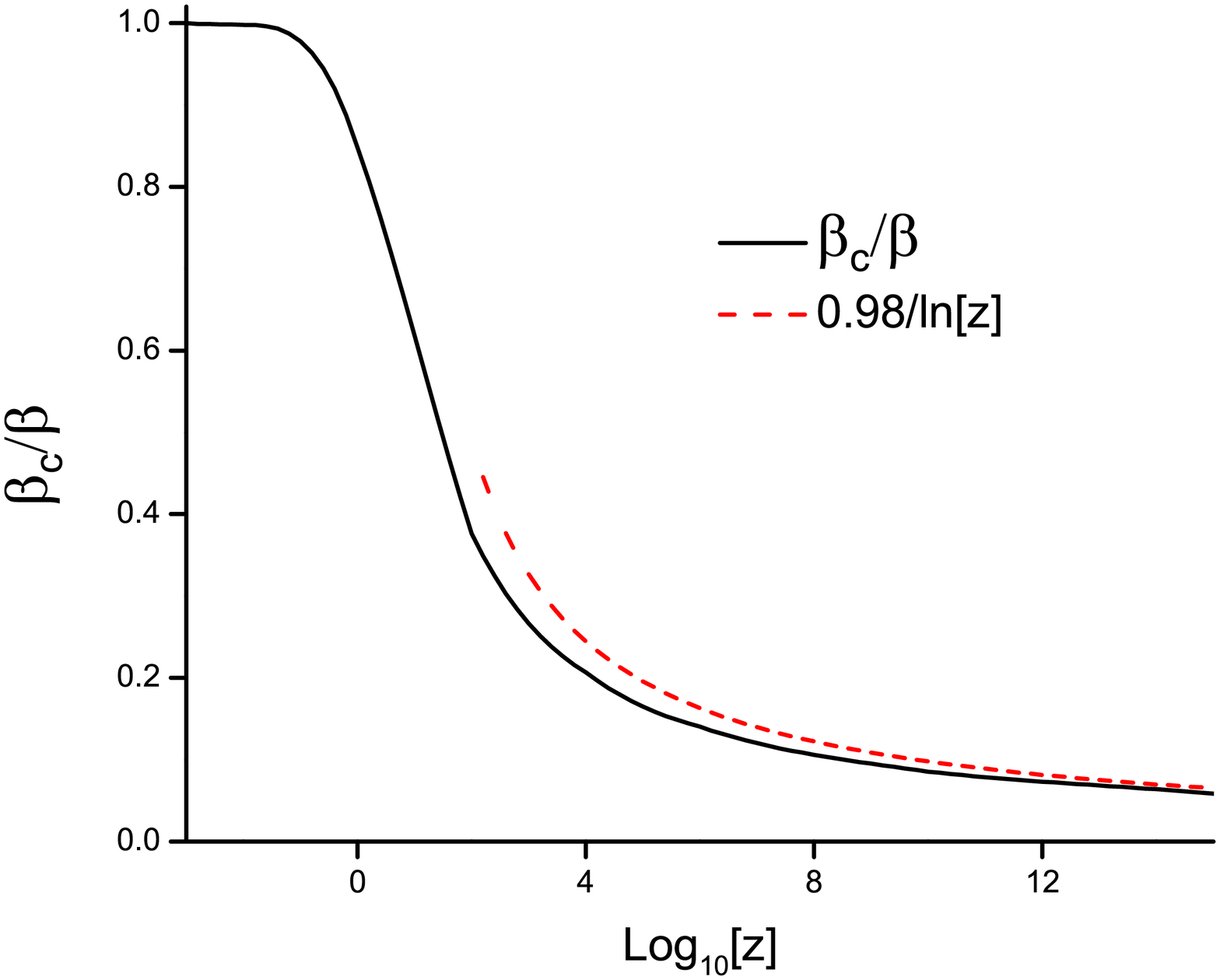}
\caption{(a) Dimensionless effective potential at various degeneracies
(z values), (b) Effective temperature as a function of z; Asymptotic limit
(dotted) implies $T_c$ is finite at $T=0$.
} \label{fig:1}
\end{figure}
Exact representations of the classical parameters as
functions/functionals of the quantum properties follow from
classical density functional theory
which has the form%
\begin{equation}
\beta _{c}\mu _{c}(\mathbf{r})=\mathcal{I}(\mathbf{r}\mid n_{c},g_{c}),%
\hspace{0.2in}\beta _{c}\phi _{c}\left( \mathbf{r},\mathbf{r}'\right)
=\mathcal{J}\left( \mathbf{r},\mathbf{r}'\mid n_{c},g_{c}\right) .  \label{3.1}
\end{equation}%
Here $\mathcal{I}(\mathbf{r}\mid \cdot ,\cdot )$ and
$\mathcal{J}\left( \mathbf{r}\mid \cdot ,\cdot \right) $ denote
formal functionals of the local
density and pair correlation function. But according to the definitions (\ref%
{2.5}) and (\ref{2.7}) these can be replaced by their corresponding
quantum properties to give the desired result
\begin{equation}
\beta _{c}\mu _{c}(\mathbf{r})=\mathcal{I}(\mathbf{r}\mid n,g),\hspace{0.2in}%
\beta _{c}\phi _{c}\left( \mathbf{r},\mathbf{r}'\right) =\mathcal{J}\left( \mathbf{r}%
,\mathbf{r}'\mid n,g\right) .  \label{3.2}
\end{equation}
This provides the formal expressions for the classical local
chemical potential and classical pair potential in terms of
properties of the quantum system, suitable for the introduction of
approximations. For example a good approximation to
$\mathcal{J}\left( \mathbf{r}\mid n,g\right) $ for Coulomb systems
is the hypernetted chain (HNC) form $\mathcal{J}\left(
\mathbf{r},\mathbf{r}'\mid n,g\right) \rightarrow -\ln (g\left(
\mathbf{r},\mathbf{r}'\right) )+g\left(
\mathbf{r},\mathbf{r}'\right) -1-c\left(
\mathbf{r},\mathbf{r}'\right)$, where $c\left(
\mathbf{r},\mathbf{r}'\right) $ is the direct correlation function,
defined in terms of $g\left( \mathbf{r},\mathbf{r}'\right) $ by the
Ornstein-Zernicke equation.

It remains to find an equivalent expression for $\beta _{c}$. This
is obtained from the first definition of (\ref{2.5}) and the
classical virial equation for the pressure, which has the form
\begin{equation}
\frac{\beta _{c}}{\beta }=\frac{\beta _{c}p_{c}}{\beta
p}=\mathcal{K}\left[
\beta _{c}\mu _{c},\beta _{c}\phi _{c},g_{c}\right] \rightarrow \mathcal{K}%
\left[ \beta _{c}\mu _{c},\beta _{c}\phi _{c},g\right] .
\label{3.4}
\end{equation}%
In the last step $g_{c}$ has been replaced by $g$ according to
(\ref{2.7}).
Since $\beta _{c}\mu _{c}$ and $\beta _{c}\phi _{c}$ are known from (\ref{3.2}), $%
\beta _{c}/\beta $ is given entirely in terms of properties of the
quantum system.

\section{Example - uniform ideal Fermi gas}
\label{sec4} An interesting first illustration of this
correspondence of classical and quantum system is the uniform ideal
Fermi gas. The quantum description of this non-interacting system
with only exchange effects is of course simple to analyze. However,
the corresponding classical system is a true many-body problem since
those same effects translate into a pair potential among all
particles \cite{Lado}. The full many-body problem must be solved to
find that effective potential. Here that problem is addressed in
terms of the approximate HNC integral equation. Only the results will be given. In Figure (\textbf{1a})
the classical pair potential is shown as a function of $r/\lambda $ where $%
\lambda $ is the thermal de Broglie wavelength defined in terms of
the quantum $\beta $. A family of curves is given for different
values of the degeneracy parameter $z=\exp \left( \beta \mu \right)
$ for the quantum system. Of interest is the development of long
range $r^{-2}$ behavior as the temperature is lowered (large z). For
this reason the mapping to a classical system requires a uniform
neutralizing background to cancel this divergent behavior.

The classical temperature is determined from the virial equation,
with $g(r)$
defined as above from the HNC integral equation. This is shown in Figure (\textbf{1b}).
As expected for $z<1,\beta _{c}/\beta \sim 1$ while $T_c$
approaches a finite value for $T=0$. It is interesting to note that
while the pressures of the classical and quantum systems are the
same by definition, the internal energies are the same as well.
However, this is true only if the internal energy is defined
thermodynamically and not mechanically.

\section{Electron gas }
\begin{figure}[<float>]
\includegraphics[width=90mm, height=70mm]{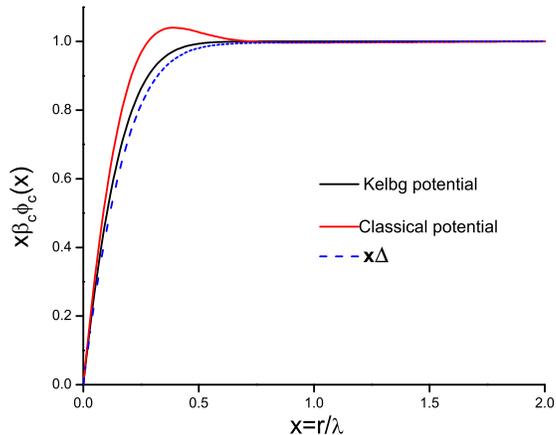}
\caption{Classical potential times $r/\lambda$ showing the regularization of Coulomb
 potential at the origin and comparison with the Kelbg potential.}
\label{fig:2}
\end{figure}
\label{sec5}A practical classical representation for the uniform
interacting electron gas can be obtained in a similar way. The
classical potential is divided into its ideal gas contribution
$\left( \beta _{c}\phi _{c}\right) ^{(0)}$ plus a contribution from
the Coulomb interactions $\Delta$: $\beta_c\phi_c=(\beta_c\phi_c)^{(0)}+\Delta$
. An approximation for the latter is
obtained by requiring the correct RPA form at weak coupling. In the
classical context this means that the direct correlation function is
given by the classical effective potential. This is obtained from
the Ornstein-Zernicke equation with the classical $g_{c}(r)$
replaced by the quantum $g(r)$, according to the definition
(\ref{2.7}), evaluated in the
RPA. The final result for the effective classical potential is%
\begin{equation}
\beta _{c}\phi _{c}\rightarrow \left( \beta _{c}\phi _{c}\right) ^{(0)}-%
\frac{1}{n}\int \frac{d\mathbf{k}}{\left( 2\pi \right) ^{3}}e^{-i\mathbf{%
k\cdot r}}\left[ \frac{S^{RPA}(k)-1}{S^{RPA}(k)}-\frac{S^{(0)}(k)-1}{%
S^{(0)}(k)}\right] .  \label{5.1}
\end{equation}%
Here $S^{(0)}(k)$ is the ideal gas static structure factor and
$S^{RPA}(k)$ is the RPA structure factor. It is possible to show
that the familiar Kelbg
potential \cite{Filinov} is recovered in the weak coupling, low density limit. \ Figure (\textbf{2})
 illustrates the quantum corrections to the bare Coulomb
form for $r_{s}=5,$ and $t=T/T_{F}=5$ (where $r_{s}$ is the usual
ion sphere radius relative to the Bohr radius, and $T_{F}$ is the
Fermi temperature).

\section{Acknowledgements}
  This research has been supported by NSF/DOE Partnership in Basic Plasma Science
and Engineering award DE-FG02-07ER54946 NSF/DOE, and by US DOE Grant
DE-SC0002139.

\end{document}